\documentclass[iop]{emulateapj}

\usepackage[colorlinks=true,linkcolor=blue,anchorcolor=blue,citecolor=blue,urlcolor=blue]{hyperref}  
\hypersetup{pdfauthor={Ho Seong HWANG}, pdftitle={A WISE View of a Nearby Supercluster A2199}}

\newcommand{\kms}{km s$^{-1}$}
\newcommand{\iras}{\textit{IRAS} }
\newcommand{\iraso}{\textit{IRAS}}
\newcommand{\akari}{\textit{AKARI} }
\newcommand{\wise}{\textit{WISE} }

\newcommand{\isoo}{\textit{ISO}}
\newcommand{\spitzer}{\textit{Spitzer} }
\newcommand{\herschel}{\textit{Herschel} }
\newcommand{\myr}{M$_\odot$ yr$^{-1}$}
\newcommand{\hmpc}{$h^{-1}$ Mpc}

\slugcomment{Last updated: \today}
\shorttitle{A \wise View of a Nearby Supercluster A2199}
\shortauthors{Hwang et al.}

\usepackage{natbib}

\begin{document}

\title{A \wise View of a Nearby Supercluster A2199}

\author{Ho Seong Hwang\altaffilmark{1}}
\author{Margaret J. Geller\altaffilmark{1}}
\author{Antonaldo Diaferio\altaffilmark{2,3}} 
\author{Kenneth J. Rines\altaffilmark{4}}

\altaffiltext{1}{Smithsonian Astrophysical Observatory, 60 Garden Street, 
  Cambridge, MA 02138, USA; hhwang@cfa.harvard.edu, mgeller@cfa.harvard.edu}
\altaffiltext{2}{Dipartimento di Fisica, 
  Universit\`a degli Studi di Torino, V. Pietro Giuria 1, 10125 Torino, Italy; diaferio@ph.unito.it,}
\altaffiltext{3}{Istituto Nazionale di Fisica Nucleare (INFN), 
  Sezione di Torino, V. Pietro Giuria 1, 10125 Torino, Italy}
\altaffiltext{4}{Department of Physics and Astronomy, 
  Western Washington University, Bellingham, WA 98225, USA; kenneth.rines@wwu.edu}



\begin{abstract}

We use Wide-field Infrared Survey Explorer ({\it WISE}) data
 covering the entire region ($\sim130$ deg$^2$) of 
 the A2199 supercluster at $z=0.03$
 to study the mid-infrared (MIR) properties of supercluster galaxies.
We identify a `MIR star-forming sequence' 
  in the \wise $[3.4]-[12]$ color-12 $\mu$m luminosity diagram,
  consisting of late-type, star-forming galaxies.
At a fixed star formation rate (SFR), 
  the MIR-detected galaxies at 22 $\mu$m or 12 $\mu$m
  tend to be more metal rich and to have higher surface brightness
  than those without MIR detection.
Using these MIR-detected galaxies, 
  we construct the IR luminosity function (LF) and 
  investigate its environmental dependence.
Both total IR (TIR) and 12 $\mu$m LFs are dominated by 
  late-type, star-forming galaxies.
The contribution of active galactic nuclei (AGN)-host galaxies
  increases with both TIR and 12 $\mu$m luminosities. 
The contribution of early-type galaxies to the 12 $\mu$m LFs increases with 
  decreasing luminosity.
The faint-end slope of the TIR LFs does not change with environment,
  but the change of faint-end slope in the 12 $\mu$m LFs with the environment
  is significant: there is a steeper faint-end slope in the cluster core 
  than in the cluster outskirts.
This steepening results primarily from 
  the increasing contribution of early-type galaxies
  toward the cluster.
These galaxies are passively evolving, and contain old stellar populations 
  with weak MIR emission from the circumstellar dust around
  asymptotic giant branch stars.
\end{abstract}


\section{Introduction}
Galaxy properties including morphology, star formation rate (SFR),
  color, and activity in galactic nuclei 
  are strongly affected by the environment (see \citealt{bm09} for a review).
The galaxy cluster environment is special
  because it contains the intracluster medium (ICM), galaxies, and dark matter
  that affect galaxy properties gravitationally and/or hydrodynamically
  over several billion years.
For example, the typical morphology (also SFR, color, and activity in galactic nuclei) 
  of galaxies changes with both local density and clustercentric radius
  (e.g., \citealt{oem74, dg76, dre80, ph09, hwa12agn}).
This environmental dependence
  of galaxy properties in galaxy clusters
  may result from a host of
  physical mechanisms including ram pressure \citep{gg72},
  cumulative galaxy-galaxy hydrodynamic/gravitational interactions \citep{ph09},
  strangulation \citep{lar80}, and galaxy harassment \citep{moo96}
  (see \citealt{bg06, ph09} for a review).
  
One of the fundamental tools for understanding the physics of
  morphological transformation and the quenching of star formation activity (SFA)
  is the galaxy luminosity function (LF; \citealt{bla01,ben03,park07,rg08,gel12}).
LFs in infrared (IR) bands can provide an unbiased view of SFA 
  in cluster galaxies
  because they are insensitive to dust extinction (e.g., \citealt{gall09,hai11}).
  
Since the pioneering work on IR observations of cluster galaxies
  with Infrared Astronomical Satellite (\iraso) and 
  Infrared Space Observatory (\isoo) (see \citealt{met05} for a review),
  there have been several determinations of the IR LFs for cluster galaxies
  with recent IR satellites including
  the \spitzer Space Telescope \citep{wer04},
  \akari \citep{mur07},
  and the \herschel Space Observatory \citep{pil10}.
For example,
  \citet{bai06,bai09} use extensive \spitzer observations of local galaxy clusters
  including Coma and A3266
  to conclude that IR LFs are well fit with the \citet{sch76} function.
They suggest that
   the bright end of IR LFs for local rich clusters has a universal form,
   similar to the LFs for nearby field galaxies.
\citet{tran09} confirmed this universal form of IR LFs in Cl1358 at $z=0.33$;
  other studies confirmed it in A2255 at $z=0.08$ \citep{shim11}
  and in the Shapley supercluster \citep{hai11}.

Studies of other clusters yield surprisingly different results : 
  there is an excess of bright IR sources in the bullet cluster at $z=0.3$ \citep{chung10}
  and an excess of faint IR sources in the A1763 supercluster at $z=0.23$ \citep{biv11}.
  
The environmental dependence of IR LFs remains an open question.
For example,
  \citet{bai09} suggest that
  the cluster and field IR LFs
  do not seem to differ significantly from one another (see also \citealt{finn10, hai11}).
However, \citet{bai06} found a hint of a steeper faint-end slope
  toward the outer region of the Coma cluster.
\citet{tran09} also found an excess of bright IR sources 
  in their LF of the super group (SG1120)
  at $z=0.37$ compared to that of field galaxies at similar redshift
  (see also \citealt{am12}).
\citet{biv11} found that the slopes of IR LFs
  in three different regions in the A1763 supercluster
  (i.e. the cluster core, the large-scale filament, and the cluster outskirts)
  are similar,
  but the filament appears to have a flatter LF than both the outskirts and the core.

Here, we discuss the IR LFs for galaxies in the A2199 supercluster.
This supercluster is one of the best targets 
  for the study of IR LFs and their environmental dependence
  because the entire supercluster region ($R\lesssim10$ \hmpc) is 
  uniformly covered by the Wide-field Infrared Survey Explorer 
  ({\it WISE}; \citealt{wri10})
  with excellent sensitivity at mid-IR (MIR) wavelengths.
This region is also fully covered by 
  the Sloan Digital Sky Survey (SDSS; \citealt{york00})
  as well as by other large spectroscopic surveys.
Thus, ambiguity in determining cluster membership is vastly reduced.
To understand the behavior of the LFs in the supercluster,
  we also explore the MIR colors of galaxies
  focusing on the connection to IR luminosities and their environmental dependence.

A2199 is a regular, X-ray bright, rich galaxy cluster at $z\sim0.03$,
  and forms a supercluster with several nearby groups in the infall region
  including A2197W and A2197E \citep{rines01}.
The center of the cluster is dominated by a massive cD galaxy, NGC 6166 \citep{kel02},
  and the cluster hosts a cooling flow \citep{john02}.

Section \ref{data} describes the observational data we use.
We construct total IR (TIR; 8$-$1000 $\mu$m) and 12 $\mu$m LFs 
  for several subsamples in \S \ref{results}.
We discuss the results and conclude 
  in \S \ref{discuss} and \S \ref{sum}, respectively.
Throughout we adopt, unless explicitly mentioned otherwise, 
  flat $\Lambda$CDM cosmological parameters of 
  $H_0 = 100~h$ km s$^{-1}$ Mpc$^{-1}$, $\Omega_{\Lambda}=0.7$ and $\Omega_{m}=0.3$.
One arcmin corresponds to $\sim$26.2 $h^{-1}$kpc at the redshift of A2199. 

\section{The Data}\label{data}

\subsection{Galaxy Catalog}

We first constructed a master catalog containing a photometric sample of galaxies
  with $m_r<20.11$ 
  (down to the magnitude where the spectroscopic samples exist)
  in the SDSS data release 7 (DR7, \citealt{aba09}).
We selected galaxies
  within 6.7\arcdeg ($\sim$10 $h^{-1}$ Mpc) of
  the A2199 center 
  ($\alpha = 16^{\rm h}28^{\rm m}38^{\rm s}$,
   $\delta = + 39^\circ 32\arcmin 55\arcsec$; \citealt{boh00}).

Spectroscopic data for galaxies with $m_r<17.77$
  are available in the SDSS database.
However, the spectroscopic completeness of the SDSS data is
  poor for bright galaxies with $m_r<14.5$ and
  for galaxies in high-density regions.
Thus, we supplement the galaxy data 
  to reduce the effects of incompleteness.
We compiled redshifts 
  for the photometric sample of galaxies in the master catalog
  from the literature
 (see \citealt{hwa10lirg} for details).
We also included the data from extensive spectroscopic survey programs
  in the field of A2199 \citep{rines02, rg08}
  which include galaxies fainter than the SDSS limit.

\begin{figure}
\center
\includegraphics[width=85mm]{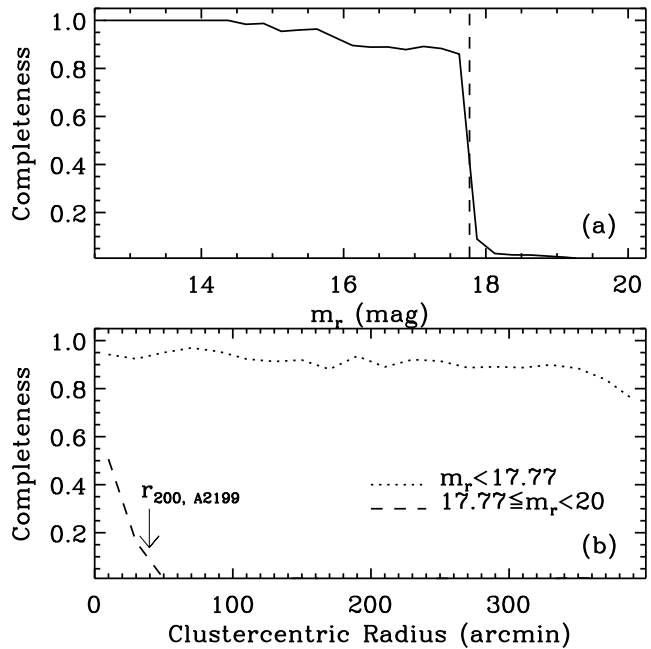}
\caption{Spectroscopic completeness of our galaxy catalog in the field of A2199
  as a function of $r$-band magnitude (a) and clustercentric radius (b). 
Vertical dashed line indicates the SDSS limit of the main galaxy sample (i.e. $m_r<17.77$).
The dotted line in (b)
  is the completeness for galaxies brighter than this limit.
The dashed line is for those fainter than the limit.
An arrow indicates $r_{200}$ of A2199.
}\label{fig-scomp}
\end{figure}

Figure \ref{fig-scomp} shows the spectroscopic completeness of the galaxy sample
  as a function of apparent magnitude and of clustercentric radius.
The spectroscopic completeness of our sample brighter than the SDSS limit
  is $\gtrsim90\%$ at all magnitudes and clustercentric radii.
Galaxies fainter than the SDSS limit 
  are not completely covered by the spectroscopic observations,
  but there are useful data within 
  the virial radius of A2199 ($r_{\rm 200,A2199}$; to be defined in \S \ref{member}).

We also use several value-added galaxy catalogs (VAGCs) drawn from SDSS data.
We adopt the photometric parameters from the SDSS pipeline \citep{sto02}.
We take the spectroscopic parameters including
  SFRs \citep{bri04} and oxygen abundance \citep{tre04} from 
  the MPA/JHU DR7 VAGC\footnote{http://www.mpa-garching.mpg.de/SDSS/DR7/}.
  
We adopt galaxy morphology data from the
  Korea Institute for Advanced Study (KIAS) DR7 VAGC\footnote{http://astro.kias.re.kr/vagc/dr7/} 
  \citep{choi10}.
In this catalog, galaxies are divided into early (ellipticals and lenticulars) and 
  late (spirals and irregulars) morphological types
  based on their locations in the ($u-r$) color versus ($g-i$) color
  gradient space and in the $i$-band concentration index space \citep{pc05}.
The resulting morphological classification
  has completeness and reliability reaching 90\%. 
We performed an additional visual check of the color images of 
  the galaxies misclassified by the automated scheme,
  and of the galaxies that are not included in the KIAS DR7 VAGC.
In this procedure,
  we revised the types of blended or merging galaxies, blue
  but elliptical-shaped galaxies, and dusty edge-on spirals.

\subsection{\wise}\label{wise}

We use the new wide-field MIR data obtained by the \wise satellite,
  which covers all the sky
  at four MIR bands (3.4, 4.6, 12 and 22 $\mu$m). 
The \wise all-sky source catalog\footnote{http://wise2.ipac.caltech.edu/docs/release/allsky/}
  contains photometric data for over 563 million objects.
\wise covers the entire region of the A2199 supercluster to a homogeneous depth and
  detects galaxies in this supercluster  
  down to L$_{\rm IR}\sim10^9$ L$_\odot$.
  
We identified \wise counterparts of the galaxies in our master catalog
  by cross-correlating them
  with the sources in the \wise All-sky data release
  with a matching tolerance of 3\arcsec($\sim$ 0.5$\times$FWHM of the PSF at 3.4 $\mu$m).
We use the point source profile-fitting magnitudes,
 and restrict our analysis to the sources with S/N$\geq3$ at each \wise band.
\wise 5$\sigma$ photometric sensitivity is estimated to be better 
  than 0.08, 0.11, 1 and 6 mJy
  at 3.4, 4.6, 12 and 22 $\mu$m
  in unconfused regions on the ecliptic plane \citep{wri10}.
 
Because 22 $\mu$m \wise data
  are closer to the peak of IR emission than other bands and 
  because they are less affected by polycyclic aromatic hydrocarbon (PAH) emission features,
  we computed the TIR luminosity ($L_{\rm IR}$) from the 22 $\mu$m flux density
  using the spectral energy distribution (SED) templates of \citet[CE01]{ce01}.
CE01 contains 105 SED templates with different TIR luminosities 
  ($3\times10^8 L_\odot <L_{\rm IR}<6\times10^{13} L_\odot$).
The templates provide $\nu L_\nu$ ($L_\odot$) as a function of wavelength.
For the observed 22 $\mu$m flux density, 
  we choose the closest template to obtain the appropriate TIR luminosity.
We interpolate between the two closest templates to reach the observed 22 $\mu$m flux density.
TIR luminosities extrapolated from a single passband have been examined in many papers.
They agree very well with those based on all far-IR (FIR) bands; 
  the uncertainty is $\sim40\%$ (e.g., \citealt{elb10, elb11}).
Therefore, this procedure does not introduce any bias in our results.
We reexamine it in \S \ref{sfr}.

\begin{figure}
\center
\includegraphics[width=85mm]{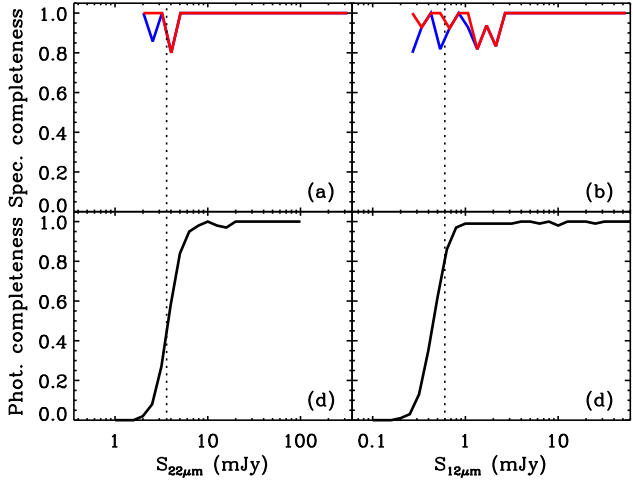}
\caption{({\it Top}) Spectroscopic completeness of 
  \wise-detected ($>3\sigma$) galaxies
   with $0.0089<z_{\rm phot}<0.0599$
  at 22 $\mu$m ({\it Left}) and at 12 $\mu$m ({\it Right}).
Blue and red curves are the completeness
  computed for the galaxies brighter than the SDSS limit
  and for all the galaxies including those fainter than the limit, respectively.
({\it Bottom}) \wise photometric completeness 
  for the A2199 supercluster region with a mean coverage depth of $\sim$21
  at 22 $\mu$m ({\it Left}) and 12 $\mu$m ({\it Right}).
Vertical dashed lines indicate
  $3\sigma$ flux limits
  (3.6 and 0.6 mJy for 22 and 12 $\mu$m, respectively).
 }\label{fig-wcomp}
\end{figure}

\begin{figure}
\center
\includegraphics[width=85mm]{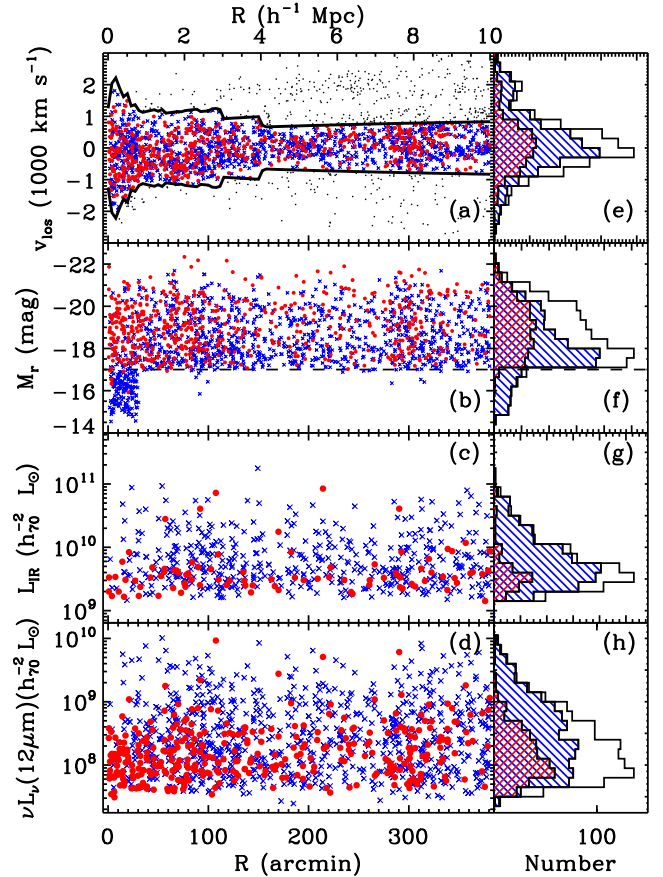}
\caption{The line-of-sight velocity relative to the cluster (a), 
  $r$-band absolute magnitude (b), 
  TIR luminosity (c) and 12 $\mu$m luminosity (d)  
  as a function of clustercentric radius, and their histograms (e-h).
We use the mean cluster redshift, $cz=9370$ \kms, determined from the caustic method.
Red filled circles and blue crosses indicate early- and late-type member galaxies, respectively.
Black dots indicate non-member galaxies.
The thick solid line indicates the estimated location of the caustics.
Open histogram show the total sample, and
  early- and late-type galaxies are denoted by 
  hatched histograms with orientation of 45$^\circ$ ($//$ with red colors) and 
  of 315$^\circ$ ($\setminus\setminus$ with blue colors), respectively.
The horizontal dashed lines in (b, f) indicate the SDSS magnitude limit.
}\label{fig-mem}
\end{figure}

\begin{figure*}
\center
\includegraphics[width=170mm]{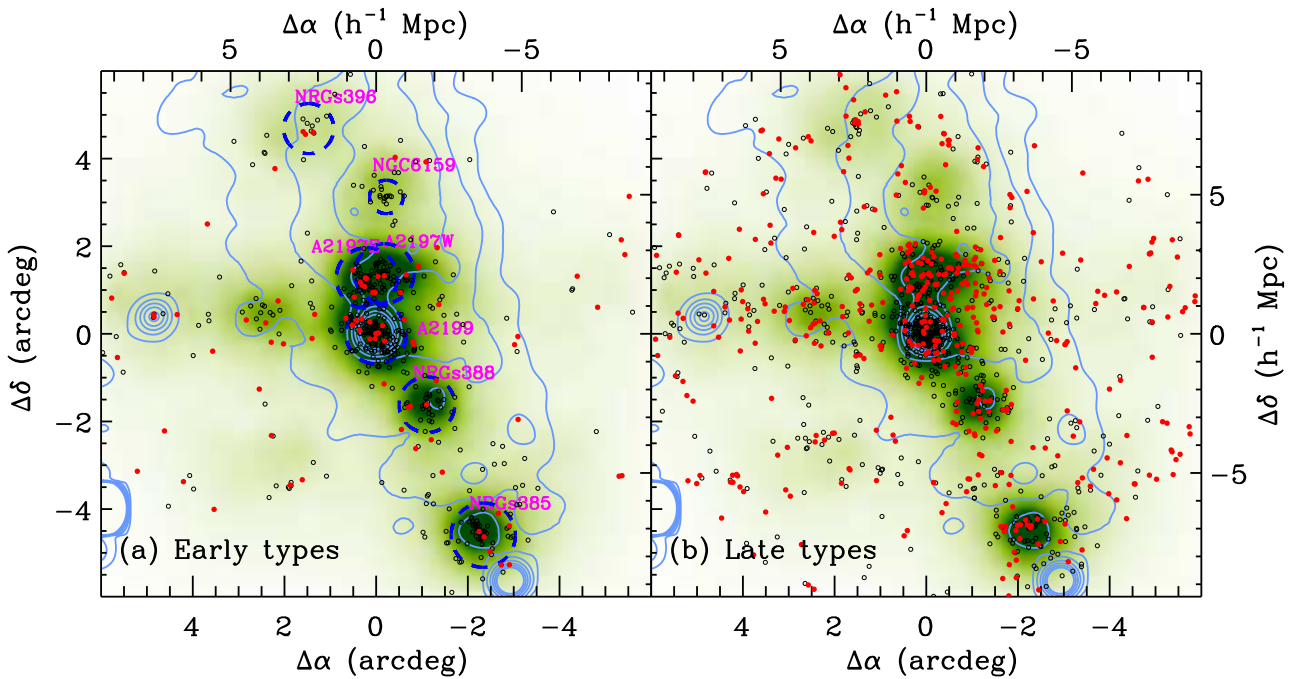}
\caption{Spatial distribution of early-type ({\it Left}) and late-type ({\it Right}) 
  galaxies in the A2199 supercluster overlaid on 
  the green and white total galaxy number density maps.
Contours with sky-blue color indicate the X-ray intensity map taken from the {\it ROSAT} All-Sky Survey,
  smoothed with a Gaussian filter of $\sigma=15$ arcmin.
Galaxies with and without \wise 22 $\mu$m detections
  are denoted by red, filled and black, open symbols, respectively.
Dashed circles in (a) represent $r_{\rm 200}$ of galaxy groups 
  in the A2199 supercluster \citep{rines01,rines02}.
North is up, and east is to the left.
}\label{fig-spat}
\end{figure*}

\subsection{Completeness}\label{compl}

To construct IR LFs for galaxies in the supercluster (\S \ref{results}),
  it is necessary to correct for the spectroscopic and photometric incompleteness 
  of our sample.
We first compute the spectroscopic completeness of \wise-detected sources 
  in the supercluster.
At the end of this section we determine the photometric completeness.
Because we are only interested in the sources in the supercluster,
  we compute the spectroscopic completeness for supercluster galaxies.
To do that,
  we select tentative member galaxies from the photometric sample
  based on the photometric redshift (photo-$z$),
  and compute the spectroscopic completeness for these galaxies.

Among several photo-$z$ measurements provided by the SDSS database,
  we adopt the photometric redshift data based on 
  the Artificial Neural Network technique \citep{oya08}.
This approach gives the tightest correlation 
  between spectroscopic and photometric redshifts 
  for galaxies at the redshift of A2199
  (see also the application to the A1763 supercluster by \citealt{biv11}).
Following \citet{kno09} and \citet{biv11},
  we determine the optimal photometric redshift range 
  for selecting tentative member galaxies
  by minimizing $\sqrt{(1-P)^2 + (1-C)^2}$.
$P$ is the purity of the sample,
  defined by the ratio of the number of 
  spectroscopically confirmed member galaxies 
  in the optimal photometric redshift range
  to the number of galaxies with any spectroscopic redshifts 
  in the optimal photometric redshift range.
$C$ is the completeness of the sample,
  the ratio of the number of spectroscopically confirmed member galaxies 
  in the optimal photometric redshift range
  to the number of spectroscopically confirmed member galaxies
  with any photometric redshifts.
From a simple experiment adjusting the redshift range,
  we found the optimal photometric redshift range for tentative members : 
  $0.0089<z_p<0.0599$.

Using these photo-$z$ selected member galaxies,
  we compute the spectroscopic completeness for the \wise-detected sources
  at 22 $\mu$m and 12 $\mu$m,
  and show the results in Figure \ref{fig-wcomp} (a-b).
Red solid lines are based on all the spectroscopic sample of galaxies
  compiled in the master catalog.
There are many galaxies fainter than the SDSS limit (i.e. $M_{\rm r}\gtrsim-17$)
  only in the central region ($R\lesssim30\arcmin$),
  but not in the outer region (see Fig. \ref{fig-mem}b).
To check any bias introduced 
  by the variation in the depth of the spectroscopic catalog with
  clustercentric radius,
  we also plot the spectroscopic completeness
  based only on the galaxies brighter than the SDSS limit (i.e. $m_r<17.77$)
  as blue solid lines.
Because the two curves are very similar,
  the LFs based on these two curves should not differ significantly
  (see \S \ref{results}).

We take the photometric completeness for \wise sources 
  from the Explanatory Supplement to the \wise all-sky data release
  products, which gives completeness curves for several 
  ($16-80$, the effective number of times that point on the sky was visited by a 
  ``good'' detector frame pixel\footnote{The exposure time for a single visit corresponds 
    to 7.7 sec (3.4 and 4.6 $\mu$m) and 8.8 sec (12 and 22 $\mu$m).}) coverage
  depth\footnote{http://wise2.ipac.caltech.edu/docs/release/allsky/expsup/sec6\_5.html}.
By interpolating the curves in the Explanatory Supplement, 
  we show the photometric completeness curves at 22 $\mu$m and 12 $\mu$m
  for the A2199 supercluster region with a coverage depth of $\sim$21 (Figure \ref{fig-wcomp}cd).

When we compute the IR LFs using {\it WISE}-detected galaxies,
  we weight each galaxy by the inverse of the photometric and spectroscopic completeness
  as a function of \wise flux density.

\begin{deluxetable}{crrcrrr}
\tablewidth{0pc} 
\tablecaption{Number of Galaxies in the A2199 supercluster
\label{tab-samp}}
\tablehead{
Wavelength & \multicolumn{2}{c}{Early types} & &  
   \multicolumn{2}{c}{Late types} \\
\cline{2-3} \cline{5-6}
($\mu$m) & AGN & non-AGN & & AGN & non-AGN & Total
}
\startdata

\hline 
\multicolumn{7}{c}{Spectroscopic sample of galaxies at $R<380$ arcmin } \nl
\hline
    All &   334  &  4322 &&  1242  &  5580  & 11478  \\
    3.4 &   332  &  4259 &&  1237  &  5381  & 11209  \\
    4.6 &   332  &  4259 &&  1231  &  5346  & 11168  \\
     12 &   289  &  2130 &&  1198  &  4741  &  8358  \\
     22 &   118  &   271 &&   892  &  2403  &  3684  \\
\hline 
\multicolumn{7}{c}{Member galaxies at $R<380$ arcmin } \nl
\hline
    All &    39  &   531 &&   135  &  1031  &  1736  \\
    3.4 &    39  &   522 &&   135  &   948  &  1644  \\
    4.6 &    39  &   522 &&   131  &   922  &  1614  \\
     12 &    37  &   309 &&   116  &   693  &  1155  \\
     22 &    16  &    70 &&    92  &   372  &   550  
           
\enddata
\end{deluxetable}

\subsection{Supercluster Membership}\label{member}

To determine the membership of galaxies in the A2199 supercluster,
  we used the caustic method \citep{dg97,dia99,serra11}.
The technique locates two curves, the caustics, 
  in the cluster redshift diagram (Figure \ref{fig-mem}a)
  that shows the line-of-sight velocities of galaxies as a function of distance 
  from the center of the supercluster. 
  
The caustics measure the escape velocity from the system of galaxies and 
  provide a basis for measurement of the mass of the system \citep{rines02}.
The caustics are also a useful tool for defining supercluster membership. 
Samples identified with the caustic technique are at least 95\% complete. 
At most 10\% of the galaxies projected within the caustics are interlopers. 
Most of these are well within the caustics (i.e. they are not velocity outliers) 
 (\citealt{serra10}; 2012, in prep.).

\begin{figure}
\center
\includegraphics[width=75mm]{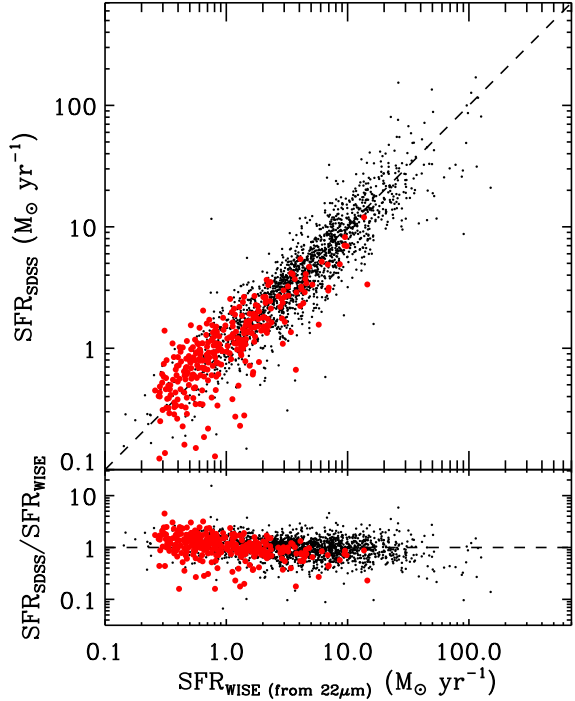}
\caption{Comparison of SFRs from \wise 22$\mu$m flux density (SFR$_{\rm WISE}$)
  with those from SDSS spectra (SFR$_{\rm SDSS}$)
We plot only galaxies with star-forming spectral type (SF, see \S\ref{agnsel}). 
Large, red circles are for member galaxies in the A2199 supercluster, and
   small, black dots are for all the galaxies at $z<0.22$ in the field of A2199.
}\label{fig-sfr}
\end{figure}

We applied the technique to a sample of 11,478 spectroscopic redshifts 
  in the field of the A2199 supercluster. 
The technique identifies 1736 members within $\sim10$ \hmpc~of the supercluster center.
Among these, 550 and 1155 galaxies are detected ($>3\sigma$)
  at \wise 22 $\mu$m and 12 $\mu$m, respectively.
The cluster center determined from this technique is consistent with
  the X-ray center used in this study (see Appendix A of \citealt{dia99} for more details).
Table \ref{tab-samp} summarizes the statistics for the number of galaxies in our sample.

In Figure \ref{fig-mem},
  we plot several physical parameters
  of the member galaxies
  as a function of clustercentric radius and morphology.
Thanks to the deep spectroscopic survey in \citet{rg08},
  many faint galaxies with $M_r>-17$ at $R<30\arcmin$
  are included as members (b).
Few luminous infrared galaxies (LIRGs, $L_{\rm IR}>10^{11}$ L$_\odot$)
  reside in this supercluster (c),
  consistent with expectation ($\lesssim 1$ LIRGs in the supurcluster volume) 
  from the field IR luminosity density at this epoch \citep{goto11lfsdss}.


Figure \ref{fig-spat} shows the spatial distribution of the member galaxies 
  segregated by their morphologies.
The galaxy number density map constructed using the member galaxies 
  with $m_r\leq17.77$ mag
  and the X-ray intensity contours from {\it ROSAT} All-Sky Survey are overlaid.

The galaxy density peaks match several known galaxy groups with X-ray emission
  shown by blue dashed circles \citep{rines01,rines02}.
The radius of the circle indicates r$_{200}$ of each group 
   (approximately the virial radius).
Within this radius, the mean overdensity is 200
  times the critical density of the universe $\rho_{\rm c}$.
We compute r$_{200}$ from the formula given by \citet{car97}:
  
\begin{equation}
r_{200}= \frac{3^{1/2}\sigma_{\rm cl}}{10 H(z)},
\label{eq-r200}
\end{equation}

\noindent where $\sigma_{\rm cl}$ is a velocity dispersion of the cluster from 
  \citet{rines02}, and  the Hubble parameter at $z$ is
  $H^2(z)=H^2_0 [\Omega_m(1+z)^3 +\Omega_k(1+z)^2+\Omega_\Lambda]$ \citep{pee93}.
$\Omega_m$, $\Omega_k$, and $\Omega_\Lambda$ are the dimensionless density parameters.
Because the observed velocity dispersions of the groups in the infall region
  could be increased by the supercluster dynamics (e.g., see Table 8 in \citealt{rines02}),
  the r$_{200}$ of each group shown in Figure \ref{fig-spat}
  should be considered only as a rough estimate of the group size.

\subsection{Comparison between WISE and SDSS data}\label{comp}

\subsubsection{Star Formation Rate}\label{sfr}

\begin{figure*}
\center
\includegraphics[width=160mm]{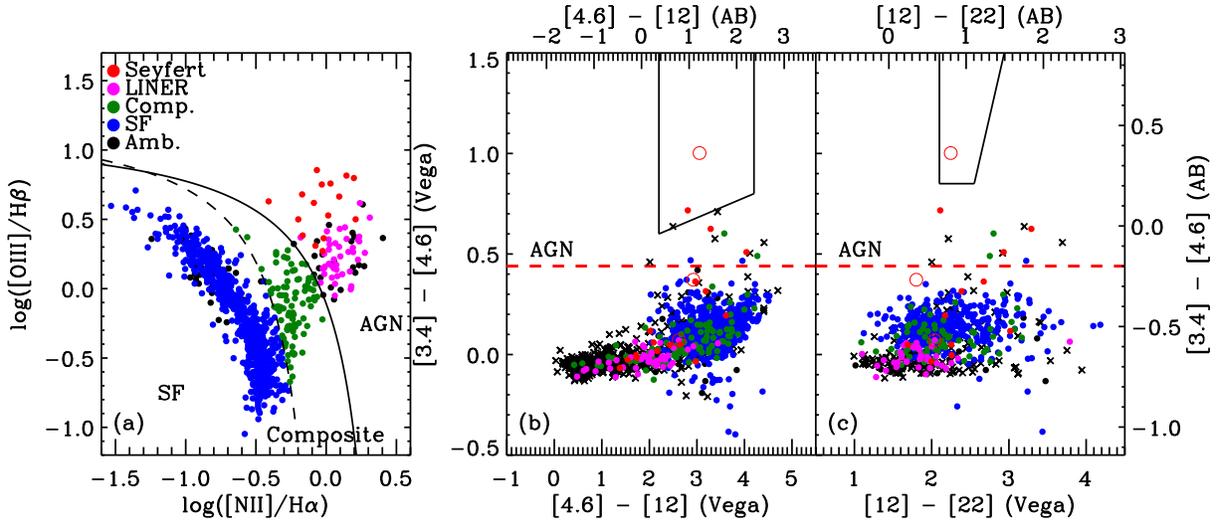}
\caption{AGN diagnostic diagrams for the member galaxies
  in the A2199 supercluster
  based on (a) optical [OIII]/H$\beta$ vs. [NII]/H$\alpha$ line ratios,
  (b-c) {\it WISE} colors. 
Different spectral types following the scheme of \citet{kew06} 
  are represented by different colored symbols
  (Seyfert: red, LINER: pink, Composite: green, SF: blue, Ambiguous: black).
The solid and dashed lines indicate the extreme starburst \citep{kew01}
  and pure SF limits \citep{kau03agn}, respectively.
Open red circles in (b-c) are Type I AGNs,
  and crosses in (b-c) are those whose 
  spectral types are not determined because of lack of optical spectra.
Solid lines in (b) and (c) are the AGN selection criteria proposed by 
  \citet{jar11} and \citet{ass10}, respectively.
Horizontal dashed lines in (b-c) are the AGN selection criteria 
  used in this study for galaxies in the A2199 supercluster.
 }\label{fig-agn}
\end{figure*}


Because we have SFR$_{\rm WISE}$ from the \wise 22 $\mu$m flux density
  and SFR$_{\rm SDSS}$ measured from the SDSS optical spectra,
  the comparison between the two measurements 
  is an important sanity check for the new \wise data
  (see also \citealt{don12}).
Among 1736 member galaxies in the A2199 supercluster,
  there are 1468 and 550 galaxies with 
    SFR$_{\rm SDSS}$ and SFR$_{\rm WISE}$, respectively.

The SFR$_{\rm SDSS}$ is from the MPA/JHU DR7 VAGC \citep{bri04},
  which provides extinction and aperture corrected
  SFR estimates of star-forming galaxies
  as well as other types of galaxies 
  (e.g., AGN, Composite, low S/N SF, low S/N LINER, and unclassifiable).
They correct the extinction
  following the dust treatment of \citet{cf00}, 
  which compares the observed line ratios with 
  those expected from models with different dust attenuations.
For those galaxies where they can not directly
  measure SFRs from the emission lines
  such as AGN and composite galaxies,
  they use the 4000-$\AA$ break (D4000) to estimate SFRs
  (see \citealt{bri04} and 
  http://www.mpa-garching.mpg.de/SDSS/DR7/sfrs.html for more details).
SFR$_{\rm WISE}$ is converted
  from the TIR luminosity
  using the relation in \citet{ken98} with the assumption of 
  a Salpeter IMF \citep{sal55}:
  SFR$_{\rm WISE}$ ($M_\odot$ yr$^{-1}$) $= 1.72\times10^{-10}L_{\rm IR} (L_\odot)$.

We show the comparison between
  SFR$_{\rm SDSS}$ and SFR$_{\rm WISE}$
  in Figure \ref{fig-sfr} based only on the galaxies with star-forming spectral type
  (see \S \ref{agnsel}).
The figure shows that SFR$_{\rm WISE}$ agrees well 
 with SFR$_{\rm SDSS}$,
  demonstrating consistency between the two measurements.

\begin{figure*}
\center
\includegraphics[width=150mm]{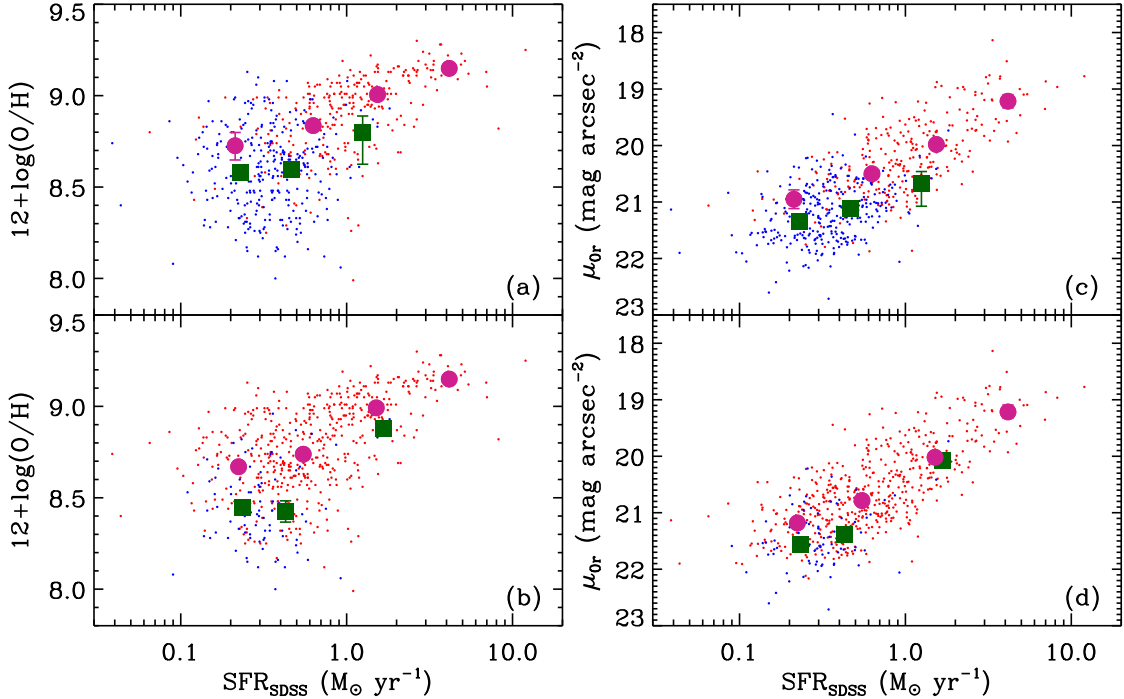}
\caption{({\it Left}) Oxygen abundance and 
  ({\it Right}) $r$-band surface brightness
  of \wise detected ($>3\sigma$) galaxies (red)
  and undetected ($<3\sigma$) galaxies (blue) 
  at 22 $\mu$m ({\it Top}) and 12 $\mu$m ({\it Bottom}).
We only plot the member galaxies in the A2199 supercluster.
Large purple circles (with \wise detection) and green squares (without \wise detection)
  are median values of each physical parameter
  and of SFR$_{\rm SDSS}$ at each bin.
The errorbars represent $68\%$ $(1\sigma)$ confidence intervals 
  determined by the bootstrap resampling method.
}\label{fig-metal}
\end{figure*}

\subsubsection{AGN Selection}\label{agnsel}

We determined the spectral types of emission-line galaxies 
  using the criteria of \citet{kew06} 
  based on the Baldwin-Phillips-Terlevich (BPT) 
  emission-line ratio diagrams \citep{bpt81,vo87}. 
For galaxies with signal-to-noise ratio (S/N)$\geq$3
   in the strong emission-lines H$\beta$, [OIII] $\lambda$5007,
   H$\alpha$, [NII] $\lambda$6584, and [SII]$\lambda\lambda$6717,6731,
   we determined the spectral types based on their positions 
   in the line ratio diagrams
   with [OIII]/H$\beta$ plotted against
   [NII]/H$\alpha$, [SII]/H$\alpha$, and [OI]/H$\alpha$.
These types are star-forming galaxies, Seyferts, 
    low-ionization nuclear emission-line regions (LINERs), 
    composite galaxies, and ambiguous galaxies 
    (see \citealt{kew06} for more details).

\begin{figure*}
\center
\includegraphics[width=170mm]{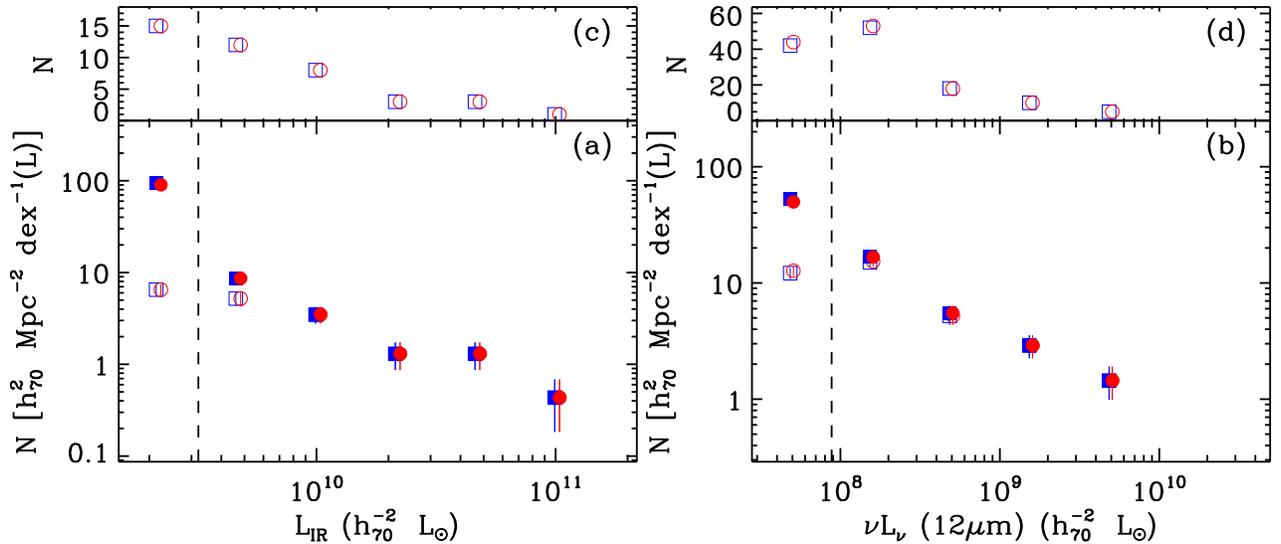}
\caption{({\it Bottom}) TIR (a) and 12 $\mu$m (b) LFs
  for galaxies in the central region of the rich cluster A2199
 (i.e. $R<r_{\rm 200, A2199} \approx 40\arcmin$).
Squares are the LFs based on the galaxies brighter than the SDSS limit.
Circles include all the galaxies fainter than the limit.
Open symbols are the raw LFs, and filled symbols are the ones after
  the corrections of photometric and spectroscopic incompleteness.
The associated errorbars indicate Poisson uncertainties.
Vertical dashed lines indicate
  the luminosity limits at the redshift of A2199
  corresponding to $3\sigma$ flux limits
  (3.6 mJy and 0.6 mJy for 22 $\mu$m and 12 $\mu$m, respectively).
({\it Top}) Number of galaxies used for the construction of the LFs 
  in each bin for TIR (c) and 12 $\mu$m (d).
Circles in the left panels are not shown because
  there are no 22 $\mu$m detections for galaxies fainter than the SDSS limit.
}\label{fig-lfa2199}
\end{figure*}

Composite galaxies host a mixture of star formation and AGN,
  and lie between the extreme starburst line \citep{kew01} and the
  pure star formation line \citep{kau03agn}
  in the [OIII]/H$\beta$ vs. [NII]/H$\alpha$ line ratio diagram 
  (see Fig. \ref{fig-agn}a).
Ambiguous galaxies are those classified as one
  type in one or two diagrams, but as another type
  in the other diagrams (see \citealt{kew06} for more details).
We assign `undetermined' type to those that do not satisfy the S/N criteria.

These AGN criteria select only Type II AGNs with narrow emission lines, 
  and miss Type I AGNs with broad Balmer lines.
To identify Type I AGNs missed in this method,
  we included galaxies with a quasar spectral classifications
  provided by the SDSS pipeline
  (i.e. \texttt{specClass} = \texttt{SPEC}$\_$\texttt{QSO} or
  \texttt{SPEC}$\_$\texttt{HIZ}$\_$\texttt{QSO}; 
  see \citealt{sto02} for more details).
Among 1736 member galaxies,
   three galaxies are Type I AGNs.

There could be still unidentified AGNs in our sample.
There are some galaxies without an SDSS spectrum
  because their redshifts are from the literature.
In some galaxies, the AGN signature may be hidden by dust (e.g., \citealt{leejc11akari}).
To identify additional AGNs,
  we plot \wise color-color diagrams in Figure \ref{fig-agn} (b-c).
Interestingly,
  the main locus of optically-selected AGN-host galaxies 
  is not clearly distinguishable 
  from the loci of other non-AGN galaxies.
If the dust is heated by AGN,
  the $[3.4]-[4.6]$ color should be red (see \citealt{ass10}).
Therefore, we use the criterion $[3.4]-[4.6]>0.44$ (Vega)
  to select AGNs at the redshift of A2199.
Among eight galaxies satisfying this criterion with known spectral types,
  six are AGN-host galaxies.
This criterion is slightly bluer than
  the AGN selection criteria used in \citet{ass10} and \citet{jar11},
  but is similar to the one in \citet{chung11}.
In summary, 
  we classify AGN-host galaxies as objects
  with either a Type I or Type II (Seyferts, LINERs and composites) AGN optical spectrum,
  and we classify MIR AGNs from the \wise color-color diagram.

\subsubsection{Metallicity and surface brightness}

To understand possible systematics in the samples of 
  \wise detected and undetected galaxies,
  we plot the oxygen abundance $[12+$log(O/H)$]$
  and the central surface brightness of
  the member galaxies in the A2199 supercluster
  as a function of SFR in Figure \ref{fig-metal}.
Following \citet{rg08},
  we compute the central surface brightness
  (in unit of mag arcsec$^{-2}$)
  from the SDSS fiber magnitudes 
  using the equation of $\mu_{0r} = m_{r,\rm fiber} + 2.123$
  assuming constant surface brightness within the fiber.

Because of increasing dust obscuration 
  (also increasing dust-to-gas ratio) with gas-phase metallicity
  (e.g., \citealt{heck98,leroy11,gmag11gdr}),
  \wise detected galaxies that have IR emission from dust
  should be more metal rich
  than undetected galaxies.
If we focus on a range 0.1 \myr $<SFR_{\rm SDSS}<$1  \myr~
  where both \wise detected and undetected galaxies exist,
  we can clearly see at both 22 and 12 $\mu$m that
  the oxygen abundance of \wise detected galaxies
  is always higher than that of undetected galaxies.
Similarly,
  the right panels show that \wise detected galaxies
  tend to have higher surface brightness than
  undetected galaxies at a fixed SFR.
Because the metallicity and the surface brightness are strongly
  coupled \citep{ryder95, simon06},
  it is difficult to conclude which is a more fundamental parameter.
In any case,
  we conclude that MIR-detected galaxies
  tend to be more metal rich and to have higher surface brightness
  than those without an MIR detection.
If we use 
  all the galaxies in the redshift range of the A2199 supercluster regardless of
  supercluster membership, the results do not change.
Thus these results are not biased by the supercluster environment.

\begin{figure*}
\center
\includegraphics[width=170mm]{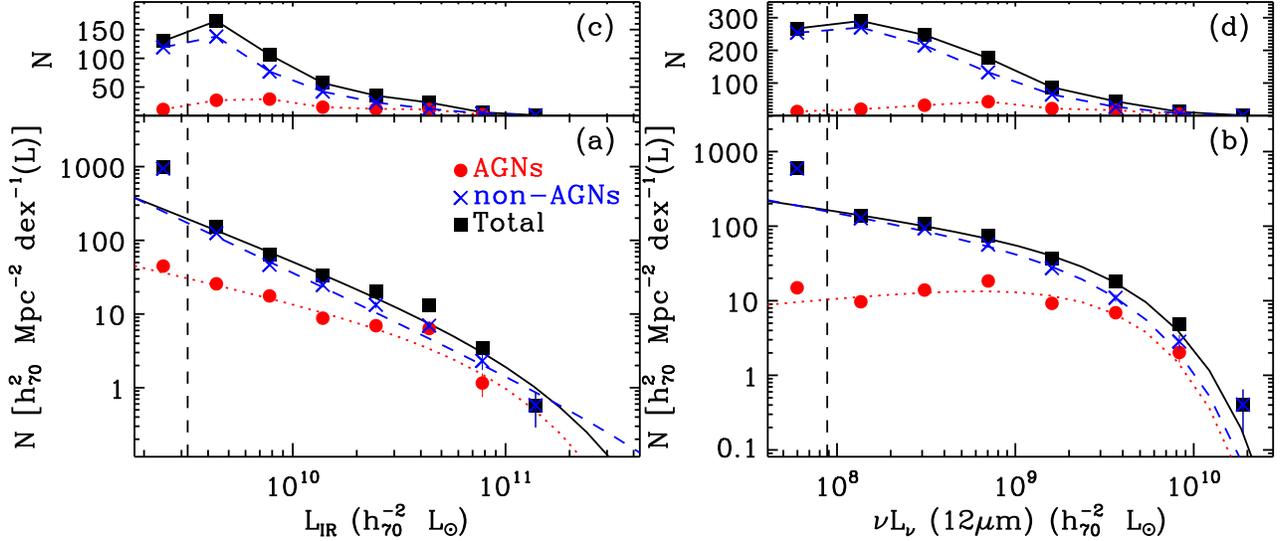}
\caption{({\it Bottom}) TIR (a) and 12 $\mu$m (b) LFs
  for galaxies in the entire supercluster region 
  (i.e. $R<380\arcmin$).
Circles and crosses are for AGN-host and non-AGN galaxies, respectively,
  and squares are for the total sample.
Dotted and dashed lines are the best-fit Schechter functions 
  for AGN-host and non-AGN galaxies, respectively,
  and solid line is for the total sample.
Vertical dashed lines indicate
  the luminosity limits at the redshift of A2199
  corresponding to $3\sigma$ flux limits
  (3.6 and 0.6 mJy for 22 and 12 $\mu$m, respectively).
({\it Top}) Number of galaxies used for the construction of the LFs 
  in each bin for TIR (c) and 12 $\mu$m (d).
 }\label{fig-lfscagn}
\end{figure*}

\begin{figure*}
\center
\includegraphics[width=170mm]{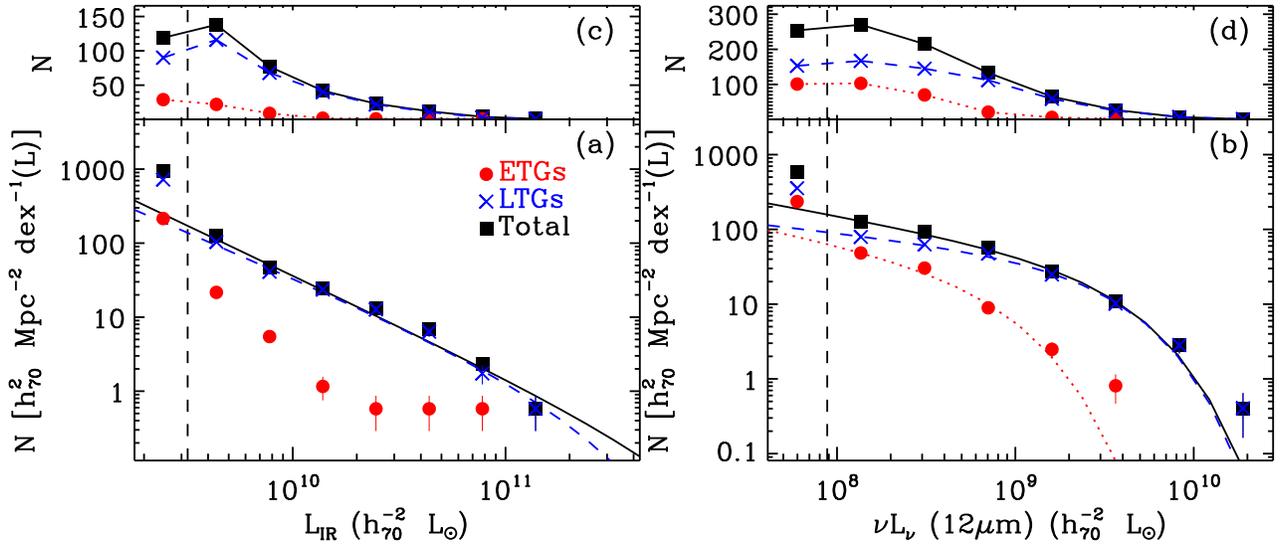}
\caption{Same as Fig. \ref{fig-lfscagn}, 
  but for early-type (ETGs, circles and dotted lines) and 
  late-type galaxies (LTGs, crosses and dashed lines) after rejecting AGN-host galaxies.
 }\label{fig-lfscmor}
\end{figure*}

\section{Luminosity Functions}\label{results}

Here we compute the projected TIR and 12 $\mu$m LFs for galaxies
  in the A2199 supercluster.
Because the depth of the spectroscopic survey
  is not the same in the central region of A2199 and in the outer regions
  (see Fig. \ref{fig-mem}b),
  we first start with the LFs in the central region
  to check for bias introduced by the variation in the depth of
  the spectroscopic survey as a function of clustercentric radius.
We construct the LFs by first counting the number of member galaxies.
We then divide by the survey area in physical size
  and by the bin size $\Delta$(log $L$).
Then we correct the counts for spectroscopic and photometric incompleteness
  by weighting each galaxy with the inverse of each completeness
  as a function of its \wise flux density (\S \ref{wise}).

Figure \ref{fig-lfa2199} shows TIR and 12 $\mu$m LFs for galaxies
  in the central cluster A2199 (i.e. $R\lesssim r_{\rm 200,A2199} \approx 40\arcmin$).
Open squares show the raw LFs without the completeness correction
  using the galaxies brighter than the SDSS limit.
The filled squares indicate the corrected LFs.
As expected, the correction only affects the faint end of the LFs.

To check the effect of the galaxies fainter than the SDSS limit
  on the derived LFs,
  we recompute the LFs again including these galaxies.
We show the results in the figure with circle symbols 
  (open for raw LFs and filled for corrected LFs).
The two corrected LFs
  based on bright (squares) and total (circles) samples agree well.
The TIR LF (left panel) based on the total sample is
  not shown because there are no 22 $\mu$m detected galaxies
  fainter than the SDSS limit in this central cluster A2199.
The agreement of two corrected LFs occurs mainly because
  the spectroscopic completeness for 
  the \wise sources in this supercluster region
  is high even when we restrict our analysis to 
  galaxies brighter than the SDSS limit (see Figure \ref{fig-wcomp}).
This result suggests that
  the corrected LFs based only on the bright sample are robust.
Because the entire region of the supercluster is covered uniformly by the SDSS data,
  we restrict the following analysis to the galaxies 
  brighter than the SDSS limit for consistency.

\begin{deluxetable*}{cccccc}
\tablewidth{0pc} 
\tablecaption{Faint-end Slopes of IR LFs in the A2199 supercluster
\label{tab-lfs}}
\tablehead{
Sample\tablenotemark{a} & Total\tablenotemark{b} & AGNs & non-AGNs & ETGs & LTGs
}
\startdata

\hline 
\multicolumn{6}{c}{TIR LFs} \nl
\hline
Entire region    &  $-2.14\pm 0.06$ & $-1.64\pm 0.13$  & $-2.36\pm 0.05$ &  ...           & $-2.23\pm0.07$ \\
Inner region   & $-2.29\pm 0.12$ & ...             & ...            &  ...           & $-2.06\pm 0.18$ \\
Outer region   & $-2.41\pm 0.11$ & ...             & ...            &  ...           & $-2.31\pm 0.12$ \\

\hline 
\multicolumn{6}{c}{12 $\mu$m LFs} \nl
\hline
Entire region &  $-1.35\pm 0.03$ & $-0.77\pm 0.08$ &  $-1.44\pm 0.04$ & $-1.48\pm0.30$ & $-1.27\pm0.05$\\
Inner region & $-1.66\pm 0.07$ & ...             & ...            & $-2.06\pm 0.25$ & $-1.38\pm 0.08$ \\
Outer region & $-1.34\pm 0.09$ & ...             & ...            & $-1.47\pm 0.31$ &  $-1.28\pm 0.10$          
     
\enddata

\tablenotetext{1}{
  Inner region : $R\leq 98\arcmin \approx 2.6$ \hmpc, Outer region : $R>249\arcmin \approx 6.5$ \hmpc.}
\tablenotetext{2}{The faint-end slopes of LFs for the total samples in the inner and outer regions,
  and for ETGs and LTGs
 are computed after rejecting AGN-host galaxies.}
\end{deluxetable*}

\begin{figure*}
\center
\includegraphics[width=170mm]{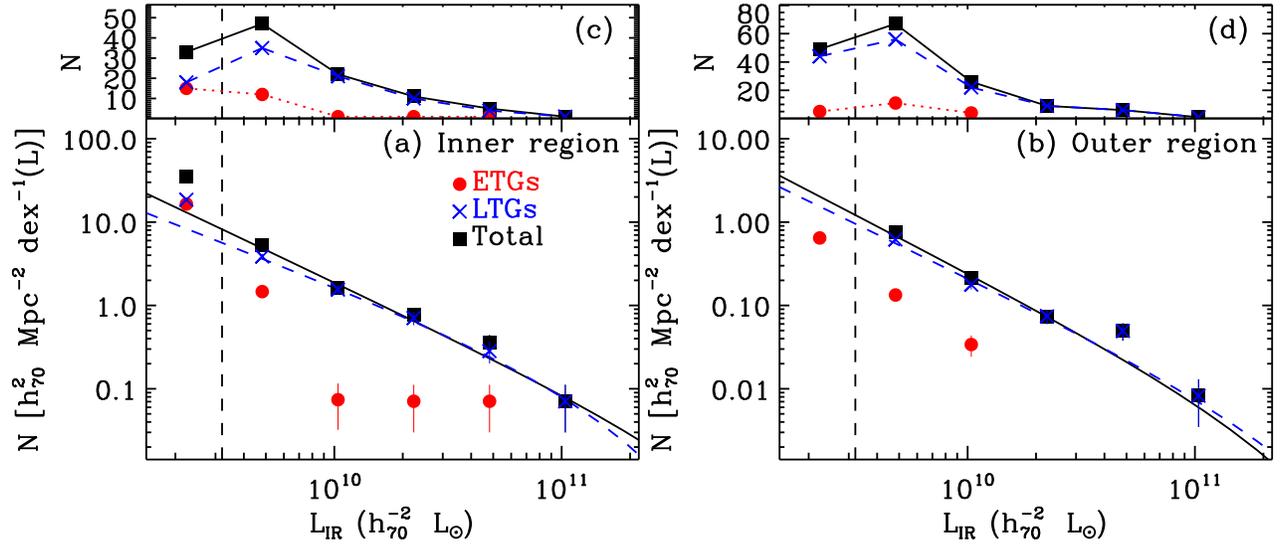}
\caption{TIR LFs for galaxies after rejecting AGN-host galaxies
  in the inner ($R\leq 98\arcmin \approx 2.6$ \hmpc) region (a) and 
  in the outer ($R>249\arcmin \approx 6.5$ \hmpc) region (b).
Circles and crosses are for early- and late-type galaxies, respectively, 
  and squares are for the total sample.
To sense the change of the slope of the LFs if any,
  we match the scales of Y-axis of the two panels.
 }\label{fig-lirlfenv}
\end{figure*}

\subsection{LFs for subsamples}

We next explore TIR and 12 $\mu$m LFs for galaxies
  in the A2199 supercluster as a whole (Figure \ref{fig-lfscagn}).
We fit the LFs with a classical \citet{sch76} function :
     
\begin{equation}
\phi (L) = \frac{dN(L)}{dA~d{\rm log}(L)} = 
  \phi^\star \left(\frac{L}{L^\star} \right)^{1+\alpha} {\rm exp} \left( - \frac{L}{L^\star} \right).
\label{eq-sch}
\end{equation}


We fit the data above the $3\sigma$ flux limit 
  (3.6 mJy and 0.6 mJy for 22 $\mu$m and 12 $\mu$m, respectively),
  shown as a vertical dashed line in each panel.
We use the MPFIT package in IDL
  \citep[an implementation of the Levenberg-Marquardt minimization]{mpfit09},
  and compute the uncertainty of the faint-end slope
  by repeating the fitting procedure 1000 times
  for random perturbations of the fitted data points within their
  errors (following a normal distribution).

Because the bright end of the LFs is not well constrained
  due to the small number of IR bright galaxies in this supercluster,
  we focus mainly on the faint-end slope of the LFs.
The faint-end slope of the TIR LF for the total sample
  is $\alpha = -2.14 \pm0.06$.
The slope of the TIR LF for AGN-host galaxies
  is $\alpha = -1.64 \pm0.13$, shallower than 
  that for non-AGN galaxies ($\alpha = -2.36\pm0.05$),
  indicating an increasing contribution of AGN-host galaxies to the TIR LFs
  with increasing IR luminosity as seen in previous studies 
  (e.g., \citealt{goto11lfsdss,goto11lfrbgs}).
Because the slopes of the LFs do not seem to change with TIR luminosity,
  we also fit the LFs with a power-law function
  (see also \citealt{biv11} for the case of the A1763 supercluster).
We obtain similar values
  of $\alpha = -2.19\pm0.03$, $-1.76\pm0.05$, and $-2.35\pm0.04$
  for the total sample, AGN-host and non-AGN galaxies, respectively.

\begin{figure*}
\center
\includegraphics[width=170mm]{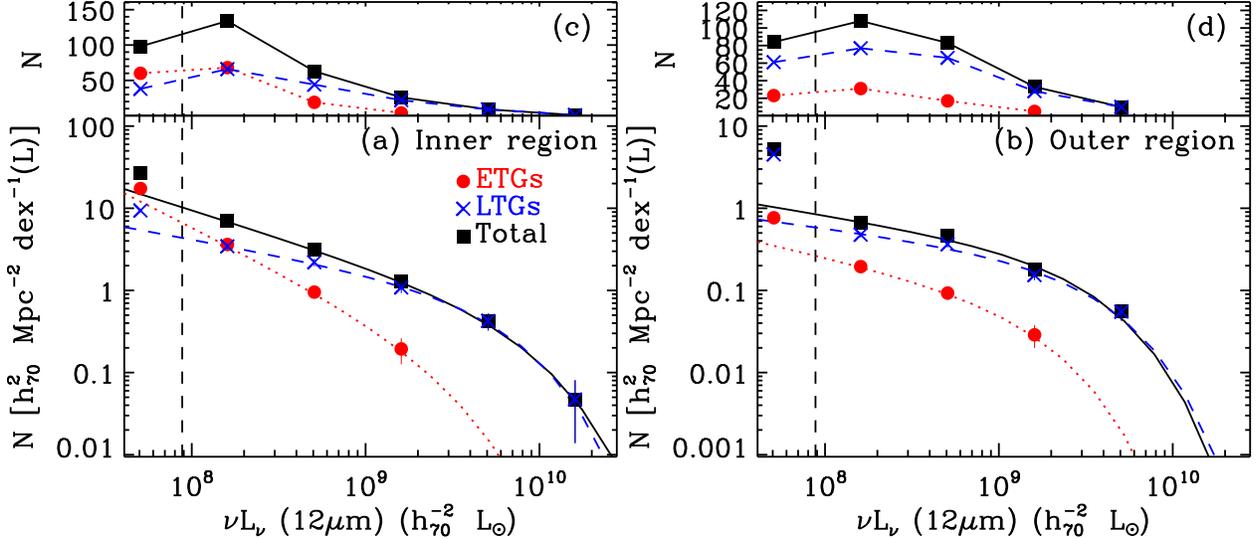}
\caption{Same as Figure \ref{fig-lirlfenv}, but for 12 $\mu$m LFs.
}\label{fig-12umlfenv}
\end{figure*}

The faint-end slope for non-AGN galaxies (i.e. SF galaxies)
  is steeper than the slope based on \iras photometry ($\alpha = -0.6$)
  for the \iras revised bright galaxy sample (RBGS; \citealt{san03})
  and the slope ($\alpha\approx -1.41$) 
  found for cluster galaxies \citep{bai06,bai09, hai11}.
The steeper slopes probably result partly from the very high spectroscopic
  completeness in this study.
However, it is only slightly steeper than
  the slope based on recent \akari photometry for RBGS
  ($\alpha = -1.9\pm0.1$ and $-1.8\pm0.1$
  for SF galaxies and for the total sample, respectively; \citealt{goto11lfrbgs}) and
  the slope for the {\it AKARI}-detected SDSS galaxies
  ($\alpha = -1.8\pm0.1$ and $-1.99\pm0.09$
  for SF galaxies and total sample, respectively; \citealt{goto11lfsdss}).
In addition, it is broadly consistent with
  the faint-end slopes
  based on all the galaxies 
  in the A1763 supercluster \citep{biv11}.

In Figure \ref{fig-lfscagn}b, 
  the AGN contribution again increases with 12 $\mu$m luminosity \citep{sm89}.
The faint-end slope is $\alpha = -1.35\pm0.03$ and $-1.44\pm0.04$
  for the total sample and for the non-AGN galaxies, respectively.
These slopes are 
  consistent with those based on \iras galaxy samples \citep{fang98}
  and those based on the local sample of {\it Spitzer}-detected galaxies \citep{pg05}.

When we decompose the LFs based on galaxy morphology 
  by rejecting AGN-host galaxies
  (see Figure \ref{fig-lfscmor}),
  the contribution of early-type galaxies to the TIR LFs is very small 
  as expected (left panels).
The faint-end slope for late-type galaxies in TIR LFs
  is $\alpha = -2.23\pm0.07$,
  again consistent with the 
  slope from the power-law function ($\alpha = -2.28\pm0.04$).
For the 12 $\mu$m LFs, 
  the faint-end slopes are $\alpha=-1.48\pm 0.30$ and $=-1.27\pm 0.05$
  for early- and late-type galaxies, respectively,
  suggesting that the contribution of early-type galaxies
  increases with decreasing luminosity
  (see the right panel of Fig. \ref{fig-lfscmor} and \S \ref{excess}).
We list the faint-end slopes for several subsamples in Table \ref{tab-lfs}.

\subsection{Environmental dependence of LFs}
\subsubsection{TIR LFs : No Environmental Dependence}

To study the environmental dependence of the LFs,
  we divide the galaxies into three radial ranges
  $R\leq 98\arcmin \approx 2.6$ \hmpc, $98\arcmin<R\leq249\arcmin$, 
  and $R> 249\arcmin \approx 6.5$ \hmpc~
  so that each range contains a similar number of galaxies.
Then we plot TIR LFs for galaxies
  only in the inner and outer regions (to emphasize the difference if any)
  in Figure \ref{fig-lirlfenv}.
We reject AGN-host galaxies in each sample.
The faint-end slope changes from
  $\alpha = -2.29\pm0.12$ (inner region) to 
  $\alpha = -2.41\pm0.11$ (outer region) for the total sample, and
  $\alpha = -2.06\pm0.18$ (inner region) to 
  $\alpha = -2.31\pm0.12$ (outer region) for late-type galaxies.
These results indicate
  no difference in the faint-end slope.
The faint-end slopes of the LFs in the intermediate region is similar.

Consistent with previous studies 
  \citep{bai09,finn10,hai11},
  our results show that the cluster and field IR LFs do not differ significantly.
\citet{biv11} also found 
  similar slopes for three different regions in the A1763 supercluster
  (i.e. the cluster core, the large-scale filament, and the cluster outskirts).
However, they also reported that 
  the filament apparently has a flatter LF than both the outskirts and the core
  (note that their LFs are complete only
  down to $L_{\rm IR}=2.5\times 10^{10}$ L$_\odot$).
Interestingly, 
  in the Coma cluster,
  \citet{bai06} suggested a hint of a steeper slope
  toward the outer region of the cluster.
Considering the large uncertainty in the determination of the faint-end slope
  (strongly affected by the spectroscopic and photometric incompleteness),
  a detailed analysis with a more comprehensive data set 
  of cluster galaxies including Coma
  is necessary to draw a strong conclusion.

Comparison of our LFs with other LFs based on different SF tracers
  is also interesting.
For example, \citet{cor05, cor08} found
  a steeper faint-end slope and a brighter $L^\star$ 
  in GALEX UV LFs for nearby clusters including
  Virgo, Coma and A1367 (also for Shapley supercluster in \citealt{hai11})
  than for the field UV LFs.
They argued that
  the steep faint-end slope observed in clusters results from
  a significant contribution of non-SF galaxies at faint UV magnitudes.
If they only consider SF galaxies,
  the cluster faint-end slope is consistent with the field.
Intriguingly, 
  the comparison of H$\alpha$ LFs between cluster (Coma, A1367 and Virgo) 
  and field galaxies
  suggests a shallower faint-end slope in clusters
  than in the field (e.g., \citealt{ip02,shi08,wes10}).
However, the comparison between the two at high-$z$ ($z\sim0.8$)
  suggests similar slopes \citep{koy10}.

\subsubsection{12 $\mu$m LFs : Strong Environmental Dependence}

Unlike the case of TIR LFs, 
  the change of faint-end slopes in 12 $\mu$m LFs
  with the clustercentric radius
  in the A2199 supercluster
  appears significant (see Figure \ref{fig-12umlfenv}):
  we find $\alpha = -1.66\pm0.07$ (inner region)
    and $\alpha = -1.34\pm0.09$ (outer region) for the total sample
  after rejecting AGN-host galaxies.
This change could result from a different morphological mix 
  and/or an intrinsically different slope of the LF 
  with clustercentric radius.
Thus we plot the LFs for early and late types separately in the figure.

\begin{figure*}
\center
\includegraphics[width=150mm]{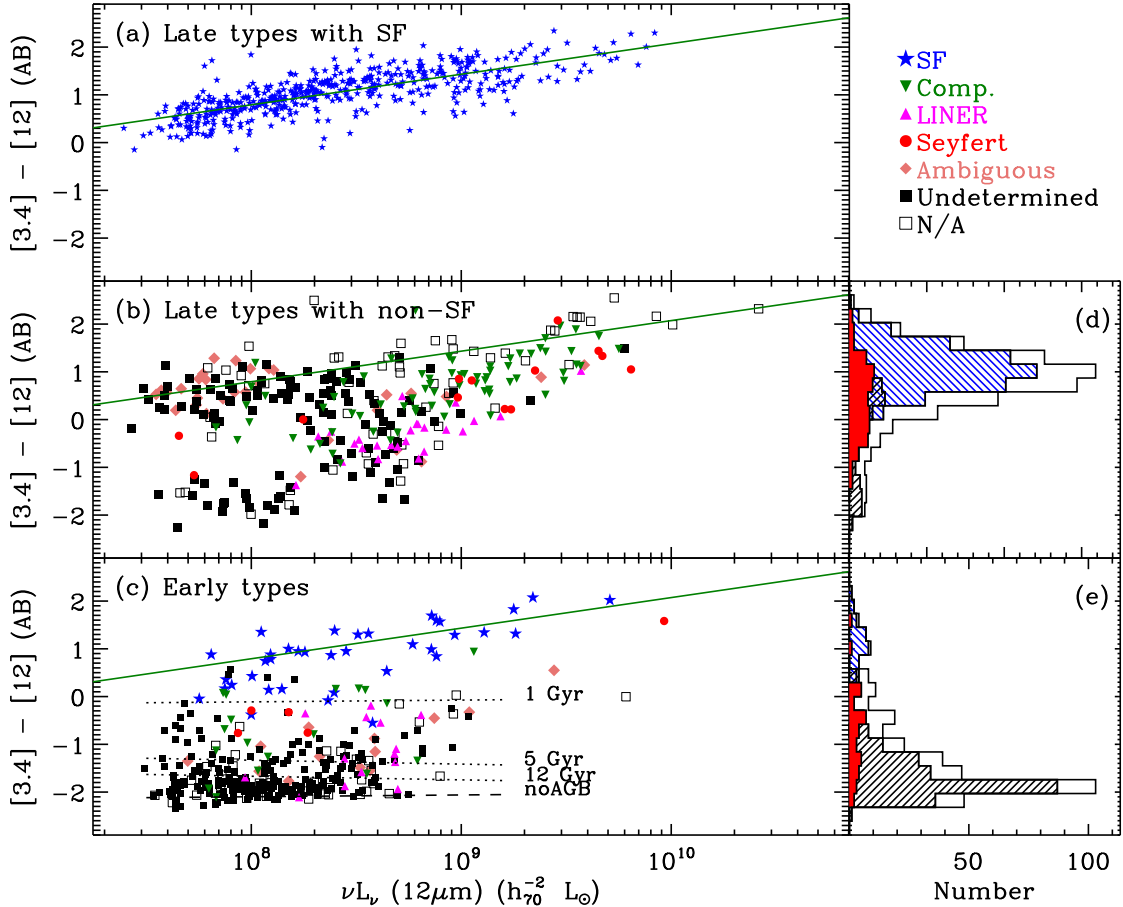}
\caption{({\it Left}) MIR color-luminosity diagram for 
  late-type galaxies with SF spectral type (a) and with non-SF spectral type (b),
  and for early-type galaxies (c) in the A2199 supercluster.
Different symbols indicate different optical spectral types.
Green solid lines are the linear fit to late-type galaxies with star formation.
Dotted lines indicate model predictions calculated 
  from the AGB model SSPs \citep{pio03}, 
  assuming a metallicity sequence at three different stellar ages 
  (1, 5, and 12 Gyr), respectively. 
The horizontal dashed line represents the model SSP without AGB dust.
({\it d}) Histograms of MIR colors for late-type, AGN-host 
  (Seyferts, LINERs, and composites) galaxies (red),
 SF galaxies [hatched one with orientation of 
   315$^\circ$ ($\setminus\setminus$ with blue colors)],
 quiescent (undetermined) galaxies 
   [hatched one with orientation of 45$^\circ$ ($//$ with black colors))], 
 and total sample (open one with black colors).
({\it e}) Same as ({\it d}), but for early-type galaxies.
 }\label{fig-mircol}
\end{figure*}

The most striking feature in Figure \ref{fig-12umlfenv} is that
  the contribution of early types to the faint-end LF 
  is comparable to that of late types
  in the inner region (a).
The faint-end slope of the LFs for early-type galaxies changes 
  from $\alpha = -2.06\pm 0.25$ (inner region) 
    to $\alpha = -1.47\pm 0.31$ (outer region), 
  and from $\alpha = -1.38\pm 0.08$ (inner region)
        to $\alpha = -1.28\pm 0.10$ (outer region)
  for late-type galaxies.

\begin{figure}
\center
\includegraphics[width=85mm]{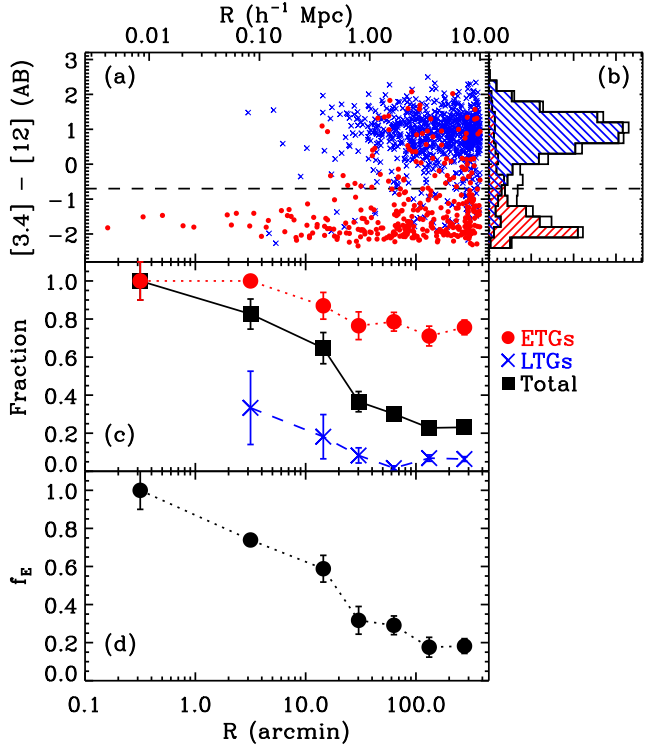}
\caption{ (a) $[3.4]-[12]$ (AB),
  (c) the fraction of weak MIR-excess galaxies  (i.e. $[3.4]-[12]<-0.7$),
  and (d) the early-type fraction ($f_{\rm E}$) 
  for galaxies in the A2199 supercluster  
  with $M_r\leq-17$
  as a function of clustercentric radius. 
(b) Histograms of MIR colors for 
  early-type [hatched one with orientation of 45$^\circ$ ($//$ with red colors)]
  and late-type galaxies
  [hatched one with orientation of 315$^\circ$ ($\setminus\setminus$ with blue colors)], 
  and total sample (open one with black colors).
Red circles and blue crosses in (a,c) are for early- and 
  late-type galaxies, respectively,
  and black squares are for the total sample.
Error bars indicate Poisson uncertainties.
}\label{fig-mirrad}
\end{figure}

These results suggest that  
  the slope of 12 $\mu$m LFs 
  changes with the clustercentric radius only for early-type galaxies,
  and therefore the change of faint-end slope for the total sample 
  results primarily from 
  the different morphological mix depending
  on the clustercentric radius 
  (i.e. the increasing contribution of early-type galaxies
  to the faint end of 12 $\mu$m LFs
  toward the cluster center).
If we divide the galaxies based on local density
  instead of clustercentric radius,
  the results for both TIR and 12 $\mu$m LFs do not change.

\section{Discussion}\label{discuss}
\subsection{MIR Star-Forming Sequence in the MIR color-luminosity diagram}

12$\mu$m LFs are very interesting because
  a variety of mechanisms in galaxies contribute to the SED 
  at this wavelength \citep{dl07} :
  SF driven dust continuum and PAH emission features \citep{smi07pah,rie09},
  hot dust component heated by AGN \citep{net07,mul11agn}, and
  dusty circumstellar envelopes of asymptotic giant branch (AGB) stars 
  (\citealt{bre98,pio03}; see also \citealt{kh10}).

To investigate the galaxy population responsible for the
  change of faint-end slopes in the 12 $\mu$m LFs with environment,
  we plot a MIR color-luminosity diagram for the supercluster member galaxies
  in Figure \ref{fig-mircol}.
MIR colors such as \wise $[3.4]-[22]$, $[4.6]-[12]$, or \akari $[3]-[11]$ 
  are useful indicators of the specific SFR and
  of the presence of intermediate age stellar populations 
  (e.g., \citealt{ko09,ko12,shim11,don12}).
Figure \ref{fig-mircol}a (top panel) clearly shows
  that there is a `MIR star-forming sequence'
  of late-type, SF galaxies ;
  there is a very good correlation between $[3.4]-[12]$ and 12 $\mu$m luminosity.
The sequence here does not extend to LIRGs because they are absent in this A2199 supercluster.
The linear fit to the data gives a relation with an rms $\sigma_{[3.4]-[12]}=0.30$,

\begin{equation}
[3.4]-[12] = {\rm log}(\nu L_\nu (12 \mu m)) \times (0.64\pm0.03) - (4.33 \pm0.21).
\label{eq-rs}
\end{equation}

The luminosities at 12 $\mu$m and at 3.4 $\mu$m
  show strong correlations with SFRs and stellar masses, respectively \citep{don12,li07}.
The SFR is also correlated with stellar mass \citep{noe07sf,elb07}.
Thus the MIR color, $[3.4]-[12]$, can be written as a function of SFR (or stellar mass).
Therefore, the color $[3.4]-[12]$ increases with 12 $\mu$m luminosity 
  as a result of the underlying correlations.
  
In Figure \ref{fig-mircol}b, 
  the distribution of most AGN-host galaxies are distinct from
  the locus of SF galaxies.
Their colors are slightly bluer than the SF sequence
  and their 12 $\mu$m luminosities are relatively high
  (i.e. $\nu L_\nu$(12 $\mu$m)$\gtrsim 3\times10^{8}$ $h^{-2}_{70}~L_{\odot}$).
There are some galaxies with `undetermined' spectral types
  in the faint end of the SF sequence   
  (i.e. $\nu L_\nu$(12 $\mu$m)$\lesssim 3\times10^{8}$ $h^{-2}_{70}~L_{\odot}$).
They are probably dusty star-forming galaxies
  without emission lines in their optical spectra because of
  strong dust extinction.

\subsection{Environmental dependence of the 12 $\mu$m LF : 
  Early-type galaxies with MIR emission}\label{excess}
  
In Figure \ref{fig-mircol}c,
  the early-type galaxies also show a `MIR star-forming sequence' consisting of SF galaxies,
  known as star-forming (or blue), early-type galaxies 
  (e.g., \citealt{fuk04,jhlee06,jhlee10beg}).
We check their optical color images and their positions 
  in the optical color-magnitude diagrams.
This inspection confirms  
  that they are indeed morphologically early-type galaxies with blue colors.
However, most early-type galaxies
  form a `MIR blue cloud' in the low luminosity regime
  (i.e. $\nu L_\nu$(12 $\mu$m)$\lesssim 5\times10^{8}$ $h^{-2}_{70}~L_{\odot}$);
  this cloud is mainly responsible for the faint-end slope of 12 $\mu$m LFs.
Most of the spectral types for these galaxies are 
  `undetermined' because there are no
  emission lines in the optical spectra,
  and they show a wide spread in MIR colors at $[3.4]-[12]<-0.7$.
Their SEDs are consistent with passively evolving, old
  stellar populations with weak MIR emission (weak MIR-excess galaxies)
  resulting from the circumstellar dust envelopes around AGB stars
  (\citealt{ko09,ko12}; see Fig. 4 in \citealt{shim11}).
These galaxies usually form a tight red sequence in the optical color-magnitude
  diagram, indicating a homogeneous population,
  but they show a wide spread in MIR colors depending on their stellar age.
In panel (c), we overplot the predictions from 
  Single Stellar Population (SSP) models with different ages,
  which include the MIR emission from the AGB dust \citep{pio03}.
Model predictions with mean stellar ages greater than $\sim5$ Gyr
  are consistent with the colors of these weak MIR-excess galaxies.

To check the radial distribution of these early-type galaxies,
  we plot $[3.4]-[12]$ colors and the fraction of these weak MIR-excess galaxies
  among 12 $\mu$m emitters
  as a function of clustercentric radius in Figure \ref{fig-mirrad}.   
Panels (a-b) show a clear $[3.4]-[12]$ color segregation depending on galaxy morphology
  at all the clustercentric radii.
There is a concentration of early-type galaxies with weak MIR emission
  in the inner region of the cluster
  ($R\lesssim100\arcmin \approx 2.5$ \hmpc~ $\approx 2.5r_{\rm200, A2199}$). 
Panel (c) shows that
  the fraction of these weak MIR-excess galaxies
  among 12 $\mu$m emitters
  starts to increase with decreasing clustercentric radius
  at $\sim100\arcmin$ both for early types and for the total sample
  (see also \citealt{hai06} based on the optical spectral analysis).
This trend is similar to the radial variation of 
  the early-type fraction shown in panel (d).
These results suggest that
  the weak MIR-excess galaxies (mostly early types)
  that are within the virial radius of A2199
  as well as at the infall region (i.e. $R\approx1-3r_{\rm 200, A2199}$), 
  are responsible for the steep faint-end slope of 12 $\mu$m LFs in
  the inner region.
  
Cluster, early-type galaxies with MIR emission were found
  in other studies (e.g., \citealt{ko09,shim11,cle11}).
For example,
  \citet{bres06} observed 17 Virgo early-type galaxies 
  with the \spitzer Infrared Spectrograph,
  and suggested that 76\% of their sample
  are passively evolving galaxies with a broad silicate feature,
  consistent with the emission from dusty
  circumstellar envelopes of mass-losing, evolved stars.
In the Coma cluster, \citet{cle09} found 
  that the majority (68\%) of the early-type galaxies
  have MIR and optical colors consistent with SSP models with dusty AGB envelopes.
Of course, there are also early-type galaxies with MIR emission
  in the field \citep{kan08,pan11}.
These studies
  provide a hint that the fraction of weak MIR-excess galaxies 
  is lower in the field than in cluster regions
  (Figure \ref{fig-mircol}b),
  but an extensive comparison with a larger sample is important.
    
  
Interestingly, the fraction of weak MIR-excess galaxies for late types 
  in panel (c) of Figure \ref{fig-mirrad},
  increases only in the very inner region of the cluster  
  ($R\lesssim20\arcmin \approx 0.5$ \hmpc~ $\approx 0.5r_{\rm200, A2199}$).
These late-type, weak MIR-excess galaxies 
  are similar to ``passive spirals'' that 
  are also observed in other clusters 
  at intermediate redshifts.
These galaxies may be progenitors of S0 galaxies 
  (e.g., \citealt{couch98, moran07}).
The existence of these late-type, weak MIR-excess galaxies 
  in the central cluster region means that their SFA is suppressed,
  but their morphologies remain late types.
This result indicates an important role for hydrodynamic processes 
  (that are not effective in changing galaxy structure)
  including ram pressure \citep{gg72}, 
  strangulation \citep{lar80},
  and galaxy-galaxy hydrodynamic interactions \citep{ph09}
  in the quenching of SFA
  in these central regions of clusters.
On the other hand,
  the physical processes related to the quenching of SFA and
  the morphological transformation of cluster galaxies
  may work over different timescales (i.e. 
  there may be faster changes in SFA than
  the morphological transformation)
  (e.g., \citealt{pog99, moran07, san09ed}).


\section{Conclusions}\label{sum}

Using the homogeneous data set of \wise and SDSS that cover
  the entire supercluster region,
  we examine the MIR properties of supercluster galaxies.
Our primary results are

\begin{enumerate}

\item The MIR colors ($[3.4]-[12]$) of late-type, star-forming galaxies
  correlate strongly with 12 $\mu$m luminosity.
These galaxies trace out a star-forming sequence
  in the MIR color-luminosity diagram.

\item MIR-detected (i.e. \wise 22 $\mu$m or 12 $\mu$m) galaxies
  tend to be more metal rich and to have higher surface brightness
  than non-MIR detections at a fixed SFR.

\end{enumerate}

Using these MIR-detected galaxies at 22 $\mu$m or 12 $\mu$m,
  we investigate the IR LFs and their environmental dependence 
  in the supercluster.
The main results are

\begin{enumerate}
  
\item The TIR LFs are dominated by 
  late-type, star-forming galaxies. 
The contribution of AGN hosts increases with increasing IR luminosity.
Similarly, late-type, non-AGN galaxies dominate 12$\mu$m LFs;
  the contribution of early-type galaxies increases with 
 decreasing 12 $\mu$m luminosity.

\item The faint-end slope of TIR LFs does not change with environment.
However, the faint-end slope in the 12 $\mu$m LFs 
  varies with the environment.
The faint-end slope in the dense inner cluster region
  is steeper than that in the less dense outer region.
This behavior results primarily from the 
  increasing contribution of early-type galaxies 
  to the faint end of 12 $\mu$m LFs
  with decreasing clustercentric radius.
These early-type galaxies
  contain passively evolving, old stellar populations 
  with weak MIR emission from AGB dust.

\end{enumerate}

The combination of the wide-field survey data set of \wise and 
  spectroscopic surveys
  covering the entire region of the A2199 supercluster
  provides a unique opportunity to
  study the MIR properties of supercluster galaxies and their environmental dependence.
A detailed view of SF and nuclear activity for these supercluster galaxies
  will be studied in a forthcoming paper (Lee et al. in prep.).
The combination of \wise all-sky survey data
  and the Hectospec Cluster Survey data (Rines et al. in prep.)
  will provide data
  for extending this study to other cluster systems.

\acknowledgments

We thank the anonymous referee for constructive comments that 
  helped us to improve the manuscript.
We thank Scott Kenyon for carefully reading the manuscript.
We also thank Gwang-Ho Lee, David Elbaz, Jongwan Ko, Minjin Kim, Jong Chul Lee, Jubee Sohn and Hyunjin Shim for useful discussion.
HSH acknowledges the Smithsonian Institution for the support
 of his post-doctoral fellowship.
The Smithsonian Institution also supports MJG's research.
AD acknowledges partial support from
  the INFN grant PD51 and the PRIN-MIUR-2008 grant \verb"2008NR3EBK_003"
``Matter-antimatter asymmetry, dark matter and dark energy in the LHC era''.
KR was funded in part by a Cottrell College Science Award from the Research Corporation.
This publication makes use of data products from the Wide-field Infrared Survey Explorer, 
which is a joint project of the University of California, Los Angeles, 
and the Jet Propulsion Laboratory/California Institute of Technology, 
funded by the National Aeronautics and Space Administration.

\bibliographystyle{apj} 
\bibliography{ref_hshwang} 
\end{document}